# Nonlinear imaging with all-dielectric metasurfaces

*Christian Schlickriede[1], Sergey Kruk[1,2], Lei Wang[2], Basudeb Sain[1],
Yuri Kivshar[2] and Thomas Zentgraf[1]*

[1] Department of Physics, Paderborn University, 33098 Paderborn, Germany

[2] Nonlinear Physics Centre, Australian National University, Canberra ACT 2601, Australia

**Abstract:**

Nonlinear metasurfaces incorporate many of the functionalities of their linear counterparts such as wavefront shaping but simultaneously they perform nonlinear optical transformations. This dual functionality leads to a rather unintuitive physical behavior which is still widely unexplored for many photonic applications. The nonlinear processes render some basic principles governing the functionality of linear metasurfaces not directly applicable, such as the superposition principle and the geometric optics approximation. On the other hand, nonlinear metasurfaces facilitate new phenomena that are not possible in the linear regime. Here, we study the imaging of objects through a dielectric nonlinear metalens. We illuminate objects by infrared light and record their generated images at the visible third-harmonic wavelengths. We revisit the classical lens theory and suggest a *generalized Gaussian lens equation* for nonlinear imaging, verified both experimentally and analytically. We also demonstrate experimentally higher-order spatial correlations facilitated by the nonlinear metalens, resulting in additional image features.

## Introduction

The image formation by a lens is one of the most basic and important optical principles. It can be easily described by the Gaussian lens equation that is a part of many introductory textbooks in optics[1]. As the linear optical component heavily relies on the basic electromagnetic principle of superposition, its functionality is easily understood in a ray optics picture. From a scientific point of view, the question rises how the image formation is affected if we violate these linear principles and replace the regular glass lens with one made of a nonlinear material. This is a novel approach, while most works on nonlinear imaging are in general either linked to experiments in which a nonlinear process is executed on the object or the image of an object.[2, 3] In contrast, we imagine a nonlinear lens, where the phase accumulation is accompanied by a nonlinear process through the lens itself (Fig. 1a-b). For this thought experiment, the fundamental wave incorporates the object information and the nonlinear image formation on the other side of this $\chi^{(3)}$-lens deviates from the linear regime. Even if it is practically possible to manufacture refractive nonlinear crystals with curved surfaces, the phase-matching conditions will massively impede the experimental realization. The intrinsic properties of metasurfaces make them attractive candidates for the realization of such a device.[4] In contrast to conventional bulky optics, the nonlinear response of subwavelength-thick metasurfaces does not rely on the phase-matching condition[5], instead, it is governed by localized geometric resonances[6-8]. Such nonlinear metasurfaces are built up as two-dimensional arrangements of subwavelength nanoresonators that have shown the potential to precisely and flexibly control beam parameters like phase, amplitude, polarization, spin, and angular momentum manipulation for the generated nonlinear light[9-17]. Recently, significant progress has been made in increasing the efficiency of nonlinear processes in metasurfaces facilitated by designs utilizing high-index dielectric nanoparticles supporting multipolar Mie resonances[18, 19], which offers a paradigm shift in nonlinear optics[7]. Several peculiar nonlinear phenomena have been reported recently, such as nonlinear holography[10, 14, 20], nonlinear optical encryption[21], image formation not captured by a lens equation[22] and refraction not covered by Snell's law[23]. Fundamental equations like the Snell's law of refraction or the geometrical-optics thin-lens equation need to be expanded or generalized for the nonlinear metasurfaces so that a more comprehensive understanding of physics may emerge.

Here, we study experimentally the nonlinear imaging of objects by nonlinear all-dielectric metalenses. The metalenses consist of silicon nanoparticles that support Mie resonances facilitating the enhancement of the nonlinear conversion efficiency for the third-harmonic process[24-27]. We show that the nonlinear formation of real THG images can be measured experimentally and solved by a generalized nonlinear lens equation. Furthermore, we can observe additional image features by higher-order spatial correlations that are based on the nonlinear process and can be tuned by certain experimental conditions.

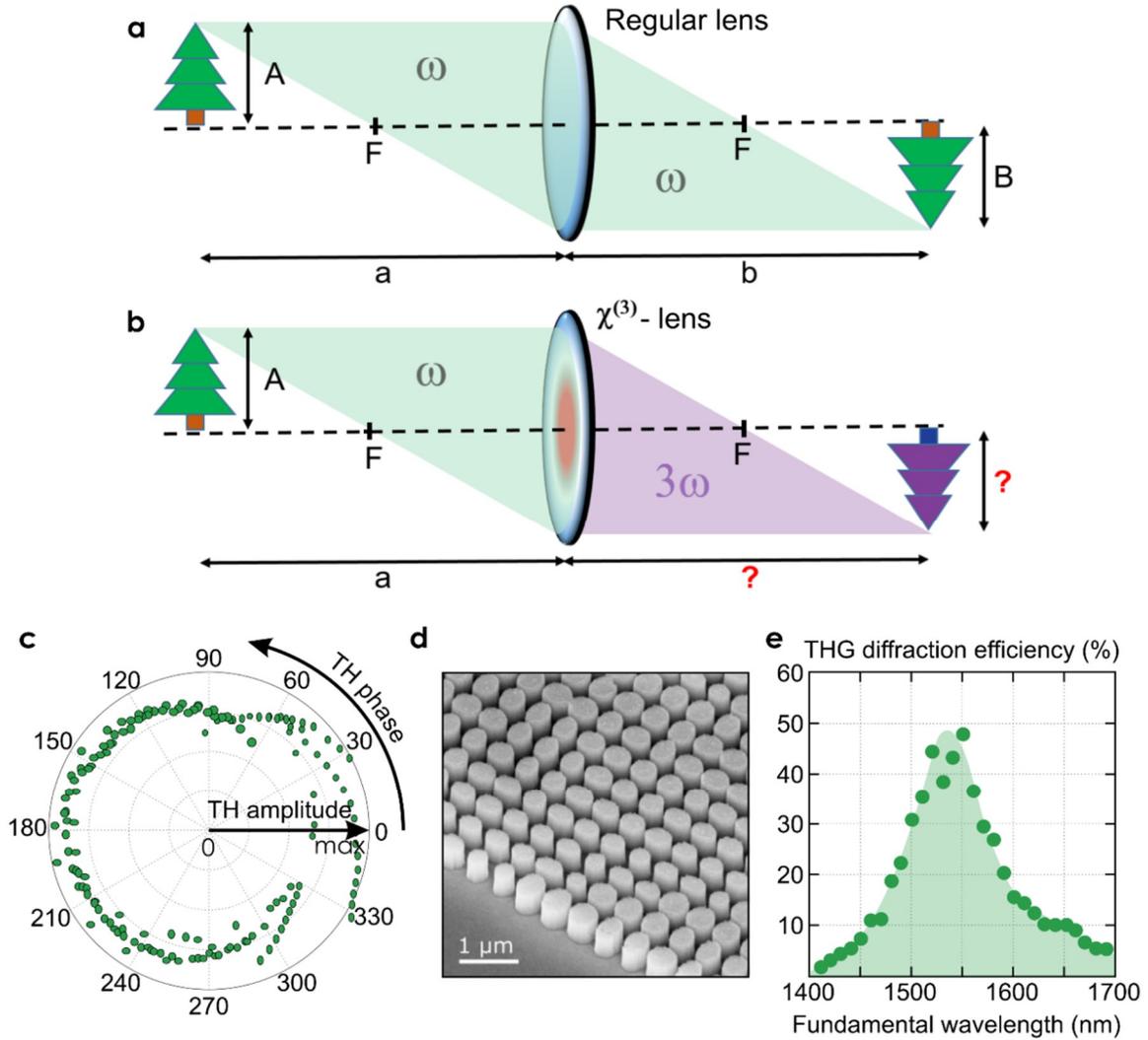

**Figure 1: Schematic concept and characteristics of nonlinear metalenses. a** Schematic illustration of the image formation with a regular lens. **b** Intuitive realization of schematic image formation with a nonlinear lens made of $\chi^{(3)}$-material. **c** Numerically simulated THG amplitude and THG phase for different nanopillar semi-axis in both x and y directions. **d** Scanning electron microscopy image of the fabricated dielectric metalens (ML400). **e** TH diffraction efficiency for corresponding fundamental illumination wavelengths.

## Results

The geometry of the nanoparticles controls the wavefront shape via the nonlinear Huygens' principle[18]. We employ a planar array of different elliptical nanopillars made from amorphous silicon (150 different nanopillar geometries in total with the full details provided in Supplementary Material). All nanopillars are of the same height of 600 nm and they are arranged into a square lattice with a period of 550 nm being placed on a glass substrate. All silicon nanopillar resonators are designed to generate a similar intensity of the nonlinear third-harmonic (TH) signal combined with the directional forward scattering. However, the different nanopillars provide different phase delays ranging from 0 to $2\pi$ via the appearing geometric resonances at the third-harmonic generation (THG) frequency. We fabricate metalenses of

200×200 µm² in size with focal distances of $f = 300$ µm (ML300) and 400 µm (ML400). The corresponding radial phase distribution for each lens is given by $\varphi(r) = \frac{2\pi}{\lambda} \left(\sqrt{f^2 + r^2} - |f|\right)$, where $r$ is the radial distance from the center of the lens, $f$ is the focal length in air and $\lambda$ is the free-space THG wavelength.

The metalenses are designed to work at the telecom wavelength of 1550 nm, which results in a TH wavelength of approximately 517 nm. Our design is experimentally verified by multispectral *k*-space analysis over the fundamental wavelengths from 1400 nm to 1700 nm (see Supplementary Material). The nonlinear metalenses reach the diffraction efficiency for the THG of 48%. The experimentally measured THG conversion efficiency is of the order of $10^{-6}$ (see Methods). In previous work[22], image formation with an appropriate plasmonic nonlinear metalens for higher harmonic generations was not demonstrated before due to the low conversion efficiency and the low damage threshold. Figure 1 c-e shows the design, electron microscopy image, and diffraction efficiency of the fabricated metalens.

Next, we study experimentally the focusing of Gaussian beams by these nonlinear metalenses. A schematic illustration is shown in Figure 2a. We trace the THG intensity distribution behind the metasurface on a camera. By splitting the beam path, we image also the intensity distribution of the pump beam in the near-infrared range on an infrared camera. The fundamental wave is loosely focused by a regular lens with a focal length of 500 mm to the metasurface so that the beam waist at the sample position is slightly larger than the size of the metalens. By changing the axial microscope objective position, we image different transverse planes on the cameras. Consequently, the propagation of the fundamental beam and TH wave can be determined along the optical axis (Fig. 2a). We find the focal Gaussian root mean square (RMS) spot sizes to be 0.5 µm for ML300 and 0.6 µm for ML400. These results are close to the respective diffraction-limited RMS spot sizes of 0.4 µm and 0.5 µm (see section S1 in Supplementary Material). Additionally, we narrow down the pump beam size (FWHM 50 µm) and illuminate only the upper part of the metalens, as shown in Fig. 2b. We trace experimentally corresponding focusing and transverse shift of the THG beam. This experiment demonstrates how the fundamental Gaussian beam is converted into the TH secondary beam: it is deflected, focused at $z = f$ to an FWHM spot size of 3.5 µm and after a propagation length of $z = 2f$ we can obtain the TH secondary beam waist of 28 µm, which is very close to the analytical prediction of $\frac{50 \mu m}{\sqrt{3}} \approx 29$ µm (see Supplementary Material).

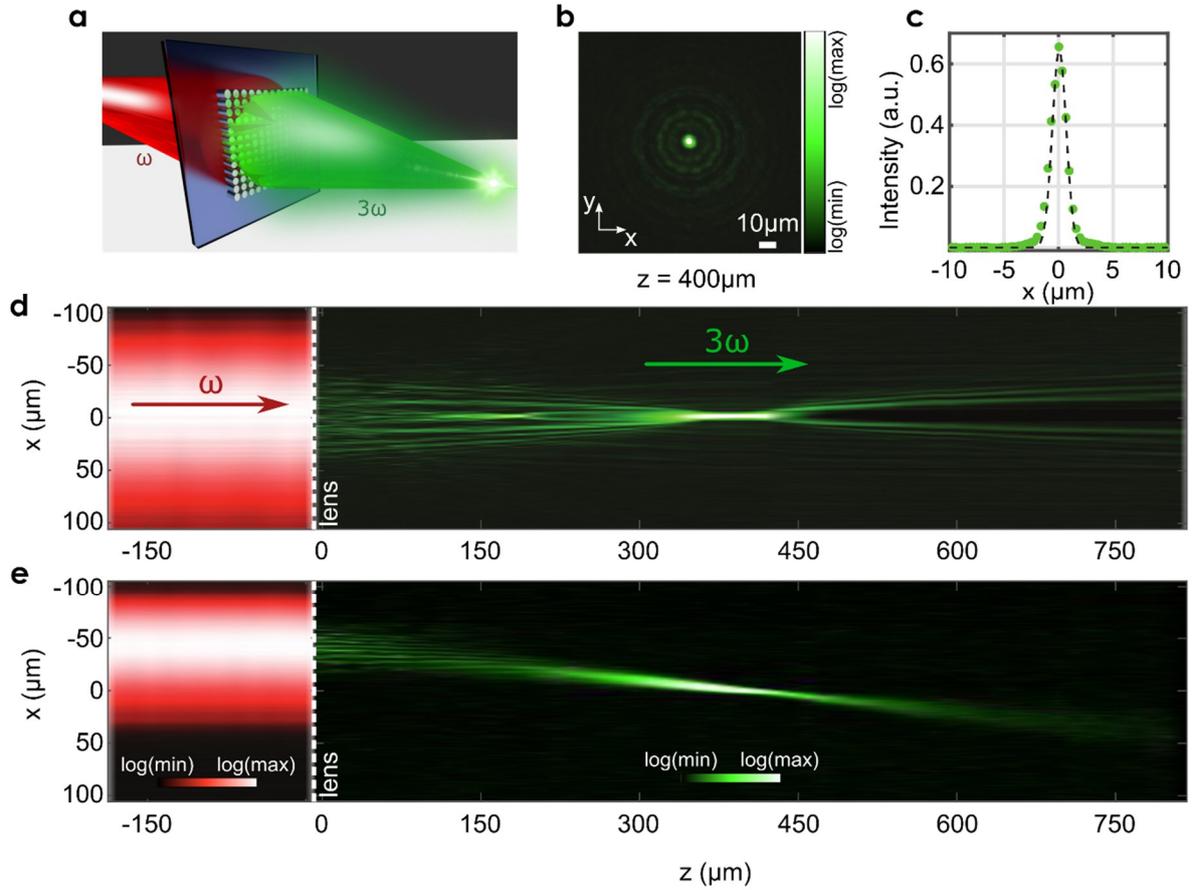

**Figure 2: Nonlinear conversion and focusing of the fundamental Gaussian beam. a** Conceptual illustration of the fundamental wave at frequency $\omega$ converted to a third-harmonic wave at frequency $3\omega$ and focused by the metasurface lens. **b** Measured TH intensity distribution of the focal spot at $z = $ **400 µm** (logarithmic scale). **c** 1D TH intensity profile showing a Gaussian RMS spot size of 0.6 µm. **d** Experimental distribution of the fundamental (red, left side) and third-harmonic (green, right side) intensities along the optical axis for the fundamental wavelength of 1550 nm and metalens focal distance of $f = 400$ µm. **e** Intensity distribution for the imaging of an off-axis Gaussian beam by the same nonlinear metalens.

Next, we proceed with the imaging of an L-shaped aperture as a real object for the TH nonlinear image formation (Fig. 3). For the imaging, we use a configuration with small $\frac{a}{l}$ parameter (where $l$ is the size of the aperture) such that the nonlinear correlation is weak and thus the nonlinear imaging process is similar to its linear counterpart. We place the aperture at an object distance of $a = -f = -$**300** µm in front of the metalens (ML300), which in linear optics should result in an image formation at infinity. Instead, we observe an image at a distance $b = $ **450** µm behind the metalens that satisfies a modified form of the Gaussian lens equation,

$$\frac{1}{f} = \frac{1}{b} - \frac{1}{n \cdot a}$$

where n is the order of nonlinear parametric process. These experimental findings are supported by theoretical calculations employing the nonlinear beam propagation method (see Fig. 3d - f). We also measured the nonlinear image formation for different object distances $a = $

($-300, -400, -500$) µm that agree with the generalized Gaussian lens equation (see Fig. 3g). Note that we use the sign convention where the object and image distances directly match the coordinates on the optical axis (positive after passing the lens and negative before passing the lens) whereas the lens is placed in the origin of the coordinate system. To reveal the physics behind this experimentally uncovered generalized lens equation, we employ numerical studies by the beam propagation method. For this, we introduce simplifications by assuming an object position close to the lens and a plane-wave illumination. This yields an expression for the generalized lens equation in the form written above (see details in Methods and Supplementary Material).

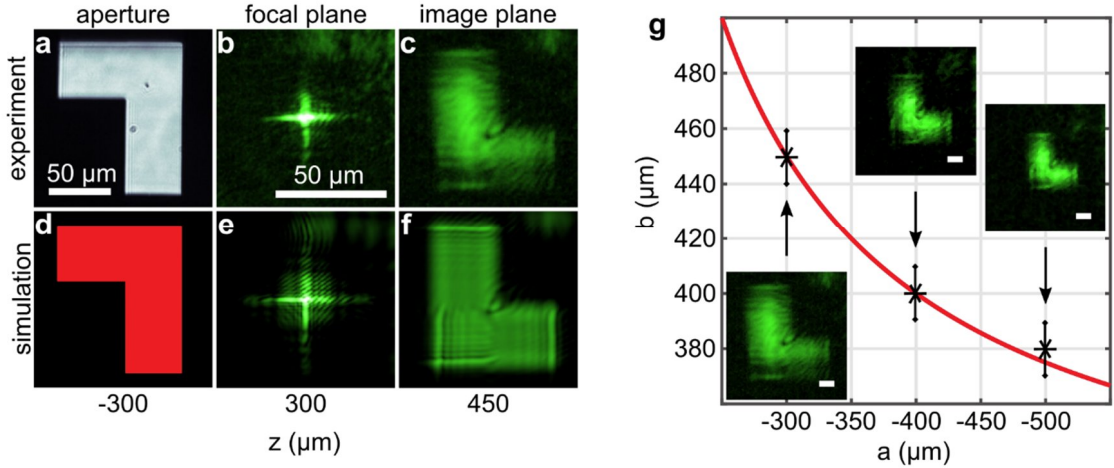

**Figure 3: Nonlinear imaging governed by the generalized Gaussian lens equation.** Upper row: **a** Microscopic white light image (transmission) of the fabricated L-shaped aperture placed at an object distance $a = -f = -300$ µm in front of the metalens (ML300). In this configuration, a conventional lens equation predicts an image formation at infinity distance. **b** Focusing of the THG at $z = 300$ µm back focal plane of the metalens. **c** TH image formation of the L-shaped aperture at an image distance of $b = 450$ µm. The real image appears inverted and demagnified in full agreement with the generalized lens equation. **d-f** Corresponding numerical simulation by using the beam propagation method on the same length scale. **g** Plot of image distances $b$ versus object distances $a$. THG image measurements are shown for different object distances $a = (-300, -400, -500)$ µm. The red curve is the prediction by the generalized Gaussian lens equation.

We further investigate the nonlinear nature of the metalens by imaging an object consisting of two apertures. The fundamental light propagates through the two apertures and diffracts during the propagation in the linear regime (see Fig. 4a, b). The signal interferes at the metalens surface and is partially converted to the third-harmonic wavelength. The added spatial phase information by the metasurface results in a focusing of the TH light. The resulting distribution of the THG signal along the optical axis is shown in Fig. 4c. Since the fundamental light reaching the metalens possesses spatial coherence, the generated nonlinear signal possesses new *spatial frequencies* leading to two additional maxima of the TH signal in the image plane (see Fig. 4d). Hence, the nonlinear image formation can, therefore, be associated with a higher-order nonlinear correlation process (see Supplementary Material, Fig. S7). This process can also be considered as a higher-order optical aberration, which only occurs in the nonlinear regime and is pronounced for relatively large object distances. We place the apertures at an object distance of $a = -3$ mm. At the location of the image plane ($z = 465$ μm), we observe four THG intensity maxima instead of two expected spots for linear optics. The resulting THG image can be understood in the following way: the larger outer circles are formed by 3ω-photons generated by ω-photons all coming from the same aperture (either top or bottom). Both inner smaller circles originate from three ω-photons generated by ω-photons from both apertures (e.g. one ω-photon from the bottom aperture and two ω-photons from the top aperture). Hence, the image formation is associated with a third-order spatial correlation function of the apertures, carrying information about spatial coherence of light coming from the object. We find that the magnitude of the higher-order correlation features at the image plane grows with the increase of the ratio $\left|\frac{a}{l}\right|$. Accordingly, these correlation features can be minimized by reducing the object distance or maximized for secondary applications that utilize higher-order correlations (like pattern recognition [28]). We support our experimental results with semi-analytical calculations using a beam propagation method that we generalize to account for the nonlinear harmonic generation (see details in Methods and Supplementary Material). The results of the corresponding theoretical calculations are shown in Fig. 4e - h.

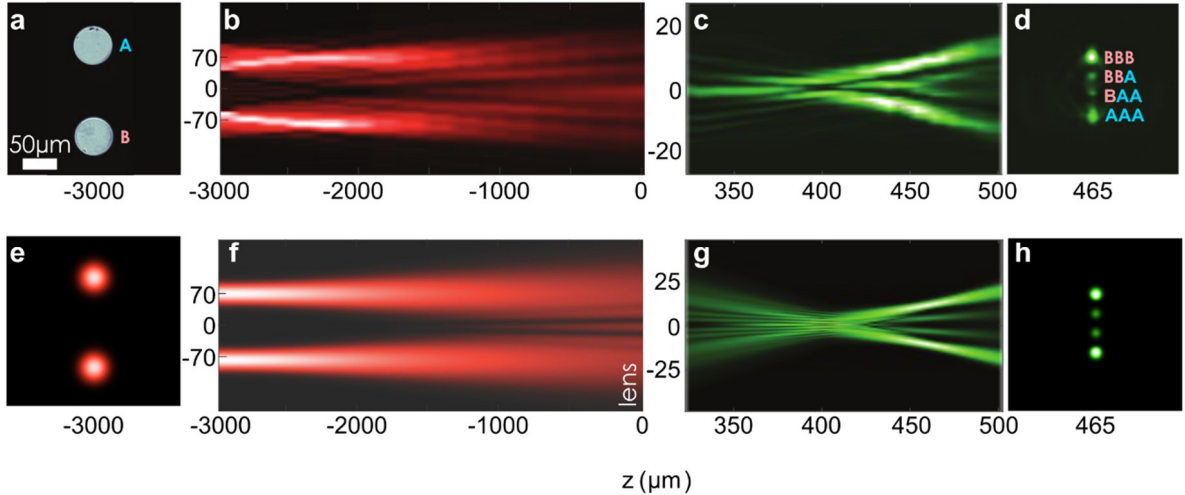

**Figure 4: Nonlinear imaging and spatial correlation for two apertures.** Upper row: **a** White light image of the two circular apertures (A and B) in front of the metalens. **b** Measured axial intensity distribution of the fundamental wave between object and metalens position. Both apertures are simultaneously illuminated with a weakly collimated beam at $\lambda = 1550$ nm. **c** Measured intensity distribution of the THG intensity along the optical axis near the focusing and imaging planes of the two apertures. **d** Cross-section of the corresponding TH intensity 465 μm after the metalens showing the formation of four maxima associated with the third-order spatial correlation function of the object. **e-h** Corresponding numerical simulation with nonlinear beam propagation method: beam waist of two Gaussian beams is located in the object plane. The fundamental and third-harmonic axial intensity distribution, as well as the image plane cross-section, is illustrated.

Next, we add a third circular aperture to the object (aperture C in Fig. 5a). At a distance of $b = 465$ μm behind the metalens, we observe a similar higher-order correlation process (Fig. 5b), which shows two additional maxima on every side of the triangular arrangement. The distance between the additional maxima on the longer side AC is greater than on the shorter sides BC and AB. Additionally, we can find one central maximum at 3ω resulting from three ω-photons coming from each of the three circular apertures. This experiment is matched with theoretical calculations using the nonlinear beam propagation method (Fig. 5c, d).

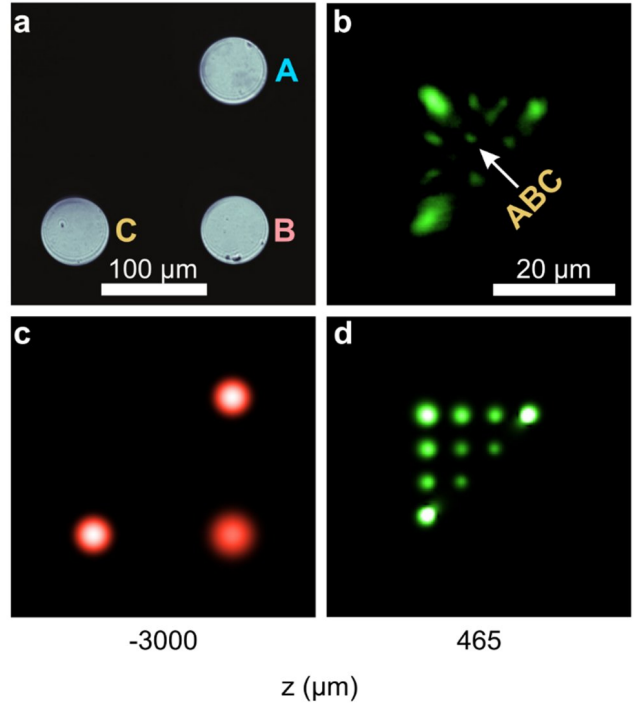

**Figure 5: Nonlinear imaging and higher-order spatial correlations for three apertures.** Upper row: **a** Microscopic image of three apertures, illuminated simultaneously at the object distance $a = -3$ mm. **b** The corresponding experimentally obtained TH intensity distribution at $b = 465$ μm behind the metalens shows the formation of 10 maxima associated with the third-order correlation. The marked spot corresponds to the THG signal obtained via nonlinear conversion of $\omega$ photons from all three apertures. **c-d** Corresponding numerical results obtained with the beam propagation method. Beam waists of three Gaussian beams are located in the object plane. The waist and the intensity of the lower right beam are chosen to be slightly different from the other two beams to reproduce the experiment more accurately.

In summary, we have demonstrated, both experimentally and theoretically, the imaging with a nonlinear all-dielectric metalens. We have employed an efficient design of a nonlinear dielectric metalens based on the generalized Huygens' principle. Our nonlinear metalens facilitates higher-order spatial correlation effects originating from the nonlinear frequency conversion not available in the linear regime. We have illustrated the conditions to maximize and minimize the additional higher-order correlation features at the image plane. Based on the experimental observations, we have formulated a generalized form of the Gaussian lens equation for imaging through general nonlinear lenses, which is in analogy to the well-known thin-lens equation in linear ray optics. The modified lens equation predicts correctly the size and location of the formed nonlinear image in our experiments. Nonlinear metalenses provide a future platform for applications in nonlinear information processing, including pattern recognition based on higher-order spatial correlations[28]; all-optical higher-order Fourier transformations; short laser pulse diagnostics; and spatial photon correlation and coherence measurements.

## Methods

*Nanofabrication.* The nonlinear all-dielectric silicon metalenses are fabricated on a glass substrate following the processes of deposition, patterning, lift-off, and etching. At first, we prepare a 600-nm-thick amorphous silicon (a-Si) film with the help of plasma-enhanced chemical vapor deposition (PECVD). As a next step, poly-methyl-methacrylate (PMMA) resist layer is spin-coated onto the a-Si film and baked on a hot plate at 170°C for 2 min. Next, the nanopillar structures are patterned by using standard electron beam lithography (EBL). The sample is then developed in 1:3 MIBK:IPA solution and washed with IPA before being coated with a 20-nm-thick chromium layer using electron beam evaporation. Afterward, a lift-off process in hot acetone was performed. Finally, by using inductively coupled plasma reactive ion etching (ICP-RIE), the structures were transferred from the chromium mask to silicon. See Ref.[22] for the fabrication of the L-shaped aperture and the circular apertures.

*Experiment.* We pump the nonlinear metalens with linearly polarized 40 femtoseconds (fs) pulses from an optical parametric amplifier (OPA). The laser repetition rate is 1 MHz and the used average power for the nonlinear measurements in Figures 1 and 5 is **~1.8 mW** (pulse energy: 1.8 nJ, peak power: 42.3 kW). Then the fs laser beam of each wavelength is focused by an achromatic condenser lens ($f_0$ = **500 mm** for the focusing experiment and the illumination of the L-shaped aperture and $f_0$ = **75 mm** for the single and multiple Gaussian beams experiments). The $f_0$ = **75 mm** condenser lens leads to a peak power density of 1.5 GW/cm² (0.3 GW/cm² for $f_0$ = **500 mm**). The metalens is illuminated at normal incidence and the THG signal is collected by an infinity-corrected microscope objective (**× 20** / 0.4**5**). A tube lens ($f_2$ = **200 mm**) is used to image the nonlinear signal onto the ultra-low noise sCMOS camera (Andor Zyla). In a second beam path, we detect the fundamental beam with an NIR camera (Xenics Xeva). Furthermore, two short-pass filters are inserted in front of the sCMOS camera to block the fundamental wave.

*Nonlinear beam propagation method.* We consider an electromagnetic field distribution $E_i(x)$ at an object distance $a$ from a lens with a focal length $f$. The output field $E_f(x)$ at a distance $b$ behind the lens in a linear regime can be calculated by the conventional beam propagation method: light propagation over an arbitrary distance $z$ is accounted by the propagator $e^{ik_z z}$ in the Fourier-space. The resulting linear beam propagation equation reads:

$$E_f(x) = \mathcal{F}^{-1}[\mathcal{F}[\mathcal{F}^{-1}[\mathcal{F}[E_i(x),k]e^{-ik_z a},x]e^{i\sigma k_0 \sqrt{f^2+x^2}},k]e^{ik_z b},x]$$

where $k_0$ is the length of the wave vector and $k_z$ accounts for the longitudinal component of the wave vector correspondingly. This is simplified by using Taylor expanded expressions corresponding to the slowly varying envelope approximation (SVEA):

$$k_x^2 + k_z^2 = k_0^2 = \left(\frac{2\pi}{\lambda}\right)^2 \quad \Rightarrow \quad k_z = \sqrt{k_0^2 - k_x^2} \approx k_0 - \frac{k_x^2}{2k_0} \quad \& \quad \sqrt{f^2+x^2} \approx f + \frac{x^2}{2f}$$

If we now take into account the beam propagation method in the nonlinear regime with harmonic generation order $n$, the electromagnetic output field changes to

$$E_f^{(nl)}(x,n) = \mathcal{F}^{-1}\left[\mathcal{F}\left[\left(\mathcal{F}^{-1}[\mathcal{F}[E_i(x),k]e^{-ik_z a},x]\right)^n e^{i\sigma n k_0 \sqrt{f^2+x^2}},k\right]e^{ink_z b},x\right]$$

where the field distribution at the lens is of the $n^{\text{th}}$ order, and the wavevector after the lens is $n$ times longer. This equation provides a general approach on how to solve for the electromagnetic output field behind the nonlinear metalens numerically, which can be easily extended to all three spatial directions. The analytical solution for the image formation, resulting in the nonlinear lens equation, can be found for particular approximations originating from large object size and object distance (see Supplementary Material).

**Acknowledgments**

We acknowledge the support from the Australia-Germany Joint Research Co-operation Scheme and the German Academic Exchange Service (DAAD). This project has received funding from the European Research Council (ERC) under the European Union's Horizon 2020 research and innovation programme (grant agreement No 724306) and the Deutsche Forschungsgemeinschaft (Grant No. DFG TRR142/C05). S.K. thanks the Alexander von Humboldt Foundation for financial support.

**References**

[1] E. Hecht. *Optics*. Pearson Education, Incorporated, 2017.

[2] Martin Oheim, Darren J Michael, Matthias Geisbauer, Dorte Madsen, and Robert H Chow. Principles of two-photon excitation fluorescence microscopy and other nonlinear imaging approaches. *Advanced drug delivery reviews*, 58(7):788–808, 2006.

[3] Christopher Barsi and Jason W Fleischer. Nonlinear abbe theory. *Nature Photonics*, 7(8):639–643, 2013.

[4] Euclides Almeida, Guy Shalem, and Yehiam Prior. Subwavelength nonlinear phase control and anomalous phase matching in plasmonic metasurfaces. *Nature communications*, 7:10367, 2016.

[5] Cheng Wang, Zhaoyi Li, Myoung-Hwan Kim, Xiao Xiong, Xi-Feng Ren, Guang-Can Guo, Nanfang Yu, and Marko Loncar. Metasurface-assisted phase-matching-free second harmonic generation in lithium niobate waveguides. *Nature communications*, 8(1):2098, 2017.

[6] Basudeb Sain, Cedrik Meier, and Thomas Zentgraf. Nonlinear optics in all-dielectric nanoantennas and metasurfaces: a review. *Advanced Photonics*, 1(2):024002, 2019.

[7] Alexander Krasnok, Mykhailo Tymchenko, and Andrea Alu. Nonlinear metasurfaces: a paradigm shift in nonlinear optics. *Materials Today*, 21(1):8–21, 2018.

[8] Guixin Li, Shuang Zhang, and Thomas Zentgraf. Nonlinear photonic metasurfaces. *Nature Reviews Materials*, 2(5):17010, 2017.

[9] Guixin Li, Shumei Chen, Nitipat Pholchai, Bernhard Reineke, Polis Wing Han Wong, Edwin Yue Bun Pun, Kok Wai Cheah, Thomas Zentgraf, and Shuang Zhang. Continuous control of the nonlinearity phase for harmonic generations. *Nature materials*, 14(6):607, 2015.

[10] Euclides Almeida, Ora Bitton, and Yehiam Prior. Nonlinear metamaterials for holography. *Nature communications*, 7:12533, 2016.

[11] Pai-Yen Chen and Andrea Alù. Optical nanoantenna arrays loaded with nonlinear materials. *Physical Review B*, 82(23):235405, 2010.

[12] Jongwon Lee, Mykhailo Tymchenko, Christos Argyropoulos, Pai-Yen Chen, Feng Lu, Frederic Demmerle, Gerhard Boehm, Markus-Christian Amann, Andrea Alu, and Mikhail A Belkin. Giant nonlinear response from plasmonic metasurfaces coupled to intersubband transitions. *Nature*, 511(7507):65, 2014.

[13] Mykhailo Tymchenko, J Sebastian Gomez-Diaz, Jongwon Lee, Nishant Nookala, Mikhail A Belkin, and Andrea Alù. Gradient nonlinear pancharatnam-berry metasurfaces. *Physical review letters*, 115(20):207403, 2015.

[14] Shay Keren-Zur, Ori Avayu, Lior Michaeli, and Tal Ellenbogen. Nonlinear beam shaping with plasmonic metasurfaces. *ACS Photonics*, 3(1):117–123, 2015.

[15] Weimin Ye, Franziska Zeuner, Xin Li, Bernhard Reineke, Shan He, Cheng-Wei Qiu, Juan Liu, Yongtian Wang, Shuang Zhang, and Thomas Zentgraf. Spin and wavelength multiplexed nonlinear metasurface holography. *Nature communications*, 7:11930, 2016.

[16] Nishant Nookala, Jongwon Lee, Mykhailo Tymchenko, J Sebastian Gomez-Diaz, Frederic Demmerle, Gerhard Boehm, Kueifu Lai, Gennady Shvets, Markus-Christian Amann, Andrea Alu, et al. Ultrathin gradient nonlinear metasurface with a giant nonlinear response. *Optica*, 3(3):283–288, 2016.

[17] Guixin Li, Lin Wu, King F Li, Shumei Chen, Christian Schlickriede, Zhengji Xu, Siya Huang, Wendi Li, Yanjun Liu, Edwin YB Pun, et al. Nonlinear metasurface for simultaneous control of spin and orbital angular momentum in second harmonic generation. *Nano letters*, 17(12):7974–7979, 2017.

[18] Lei Wang, Sergey S Kruk, Kirill L Koshelev, Ivan I Kravchenko, Barry Luther-Davies, and Yuri S Kivshar. Nonlinear wavefront control with all-dielectric metasurfaces. *Nano letters*, 2018.

[19] L Carletti, A Locatelli, O Stepanenko, G Leo, and C De Angelis. Enhanced second-harmonic generation from magnetic resonance in algaas nanoantennas. *Optics express*, 23(20):26544–26550, 2015.

[20] Bernhard Reineke, Basudeb Sain, Ruizhe Zhao, Luca Carletti, Bingyi Liu, Lingling Huang, Costantino De Angelis, and Thomas Zentgraf. Silicon metasurfaces for third harmonic geometric phase manipulation and multiplexed holography. *Nano letters*, 19(9):6585–6591, 2019.

[21] Felicitas Walter, Guixin Li, Cedrik Meier, Shuang Zhang, and Thomas Zentgraf. Ultrathin nonlinear metasurface for optical image encoding. *Nano letters*, 17(5):3171–3175, 2017.

[22] Christian Schlickriede, Naomi Waterman, Bernhard Reineke, Philip Georgi, Guixin Li, Shuang Zhang, and Thomas Zentgraf. Imaging through nonlinear metalens using second harmonic generation. *Advanced Materials*, 30(8):1703843, 2018.

[23] Jing Zhang, Xiaohui Zhao, Yuanlin Zheng, and Xianfeng Chen. Generalized nonlinear snell's law at χ (2) modulated nonlinear metasurfaces: anomalous nonlinear refraction and reflection. *Optics letters*, 44(2):431–434, 2019.

[24] Daria Smirnova and Yuri S Kivshar. Multipolar nonlinear nanophotonics. *Optica*, 3(11):1241–1255, 2016.

[25] Maxim R Shcherbakov, Dragomir N Neshev, Ben Hopkins, Alexander S Shorokhov, Isabelle Staude, Elizaveta V Melik-Gaykazyan, Manuel Decker, Alexander A Ezhov, Andrey E Miroshnichenko, Igal Brener, et al. Enhanced third-harmonic generation in silicon nanoparticles driven by magnetic response. *Nano letters*, 14(11):6488–6492, 2014.

[26] Yuanmu Yang, Wenyi Wang, Abdelaziz Boulesbaa, Ivan I Kravchenko, Dayrl P Briggs, Alexander Puretzky, David Geohegan, and Jason Valentine. Nonlinear fano-resonant dielectric metasurfaces. *Nano letters*, 15(11):7388–7393, 2015.

[27] Gustavo Grinblat, Yi Li, Michael P Nielsen, Rupert F Oulton, and Stefan A Maier. Enhanced third harmonic generation in single germanium nanodisks excited at the anapole mode. *Nano letters*, 16(7):4635–4640, 2016.

[28] John A McLaughlin and Josef Raviv. Nth-order autocorrelations in pattern recognition. *Information and Control*, 12(2):121–142, 1968.